\documentclass[sigconf]{acmart}

\usepackage{booktabs} 
\usepackage{graphicx}

\usepackage[english]{babel}

\usepackage{subfigure}
\usepackage{stfloats}
\usepackage{color}
\usepackage{dsfont}
\usepackage{amssymb}

\usepackage{bm}
\usepackage{ragged2e}
\usepackage{stfloats}
\usepackage{eucal}
\usepackage{extarrows}
\usepackage{stfloats}
\usepackage{pifont}
\usepackage{eucal}
\usepackage{multirow}
\usepackage{array}
\usepackage{extarrows}
\usepackage{enumitem}
\usepackage{epstopdf}
\usepackage{algorithm}
\usepackage{algorithmic}

\renewcommand\footnotetextcopyrightpermission[1]{} 
\begin{document}
\title{Reconstruct the Directories for In-Memory File Systems}

\author{\large Runyu Zhang, Chaoshu Yang\\ College of Computer Science, Chongqing University, China\\ \{zry.cqu, yangchaoshu\}@gmail.com}
\renewcommand{\shortauthors}{}

\begin{abstract}
Existing path lookup routines in file systems need to construct an auxiliary index in memory or traverse the dentries of the directory file sequentially, which brings either heavy writes or large timing cost. This paper designs a novel path lookup mechanism, Content-Indexed Browsing (CIB), for file systems on persistent memory, in which the structure of directory files is an exclusive index that can be searched in $O(log(n))$ time. We implement CIB in a real persistent memory file system, PMFS, denoted by CIB-PMFS. Comprehensive evaluations show that CIB can achieve times of performance improvement over the conventional lookup schemes in PMFS, and brings 20.4\% improvement on the overall performance of PMFS. Furthermore, CIB reduces the writes on persistent memory by orders of magnitude comparing with existing extra index schemes.
\end{abstract}


\maketitle

\section{Introduction}

\begin{figure}[ht]
    \centering
    \includegraphics[width=2.8 in,height = 2.1 in]{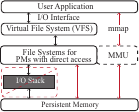}
    \caption{Direct access on PMs.}
    \label{fig:overview}
\end{figure}

Emerging persistent memory (PM) technologies \cite{3DX,burr2010phase,PCMs} provide persistent storage to complement DRAM in two aspects.
First, PMs offer significant speedup over disks and flashes.
Second, as illustrated in Figure \ref{fig:overview}, PMs can be directly accessed via load/store instructions and thus bypass the deep software stack in OS.
In recent years, many novel designs of data storage infrastructures (e.g., database and file systems) achieve higher performance and less write amplification over traditional designs by leveraging the byte-addressability of PM
 \cite{condit2009better,hu2013software,coburn2012nv,chen2011rethinking,chen2015persistent,yang2015nv,wu2011scmfs,xu2016nova,dulloor2014system}.

In traditional file systems, as the number of files growing in a certain directory, the latency of path lookup routine linearly scales up.
This is because the dentries in directory files are stored out-of-order and sequential lookup is required for searching a certain dentry.
To accelerate the lookup routine, Virtual File System (VFS) builds a hashing cache for the recently visited path.
However, for streaming applications, where most files are accessed only once, the cache mechanism fails to improve performance.
Moreover, since page cache is bypassed in PM access, such hashing cache is obsoleted.

Nevertheless, we observe that the designs for directory files of PM-oriented file systems mainly follow to that of traditional file systems on block devices.
Due to the inefficient lookup routine, these schemes are incompetent to fully exploit the potential of PMs.
To address this issue, SanGuo \cite{zeng2017efficient} devises extra indexes in DRAM for each directory file.
Indexes with tree-based structure provide steady performance for directory operations, but it needs reconstruction after power failures.
On the other hand, maintaining tree indexes persistently (like BtrFS \cite{rodeh2013btrfs}) brings extra overhead and performance penalties.
To make matters worse, modifying tree structure imports a large number of write activities on PM.
This observation implicates that the conventional directory designs degrade either performance or PM-endurance.

In this paper, we propose a novel path lookup mechanism, Content-Indexed Browsing (CIB).
To speed up the dentry queries, we replace the conventional structure of directory files by a novel content aware index.
We also reconstruct the dentries and redefine the whole lookup routine.
With the consideration of diverse access pattern between directory and data files, we redesign the operations for directory files.
By incorporating the characteristics of indexes and PM-based file systems, we can achieve optimal performance with minimum PM writes.
Specifically, the dentries in directory files can be searched in $O(log(n))$ time without building extra indexes.

We implement the proposed CIB scheme in PMFS \cite{dulloor2014system} (denoted by CIB-PMFS), a real persistent memory file system.
Then, we evaluate the performance and write activities of CIB via the widely-used benchmark Filebench \cite{tarasov2016filebench}.
Extensive experimental results show that CIB achieves 5.45$\times$ to 176.08$\times$ performance improvement over the lookup scheme in PMFS for different types of workloads, which brings 8\% to 20.4\% improvement on the overall performance.
Furthermore, CIB reduces PM writes by orders of magnitude compared with extra index schemes.

Our main contributions are listed as follows.
\begin{itemize}[noitemsep,topsep=0pt,parsep=0pt,partopsep=0pt]
  \item We conduct an in-depth investigation of the drawbacks in structure and lookup routine of existing file systems.
  \item We propose a novel Content-Indexed Browsing (CIB) scheme and redefine the structures and routines in terms of directory files.
  \item We implement our scheme in PMFS, denoted by CIB-PMFS. Substantial experiments show that CIB-PMFS can improve performance and reduce writes on PM in comparison with the original PMFS.
\end{itemize}

The rest of this paper is organized as follows.
In Section~\ref{Sec:Motivation}, we introduce the existing lookup schemes of existing file systems and show a motivational example with experiments.
We present the design and implementation of the CIB scheme in Section~\ref{Sec:DesignFramework} and Section~\ref{Sec:implementation}, respectively.
We show the evaluation results of CIB in Section~\ref{Sec:Exp}.
Section~\ref{Sec:Con} concludes the paper.

\section{Background and Motivation} \label{Sec:Motivation}

With the continuous progress of researches in Persistent Memory (PM) technologies, advanced in-memory systems are developed in succession \cite{tan2015memory,arulraj2018bztree,burr2008overview}.
Compared with block devices, there are several advantages of PMs.
First, PMs have orders of magnitude less read/write latency than block devices;
Second, PMs connect to CPU using load/store instructions rather than deep I/O stack, reducing overhead caused by storage hierarchy \cite{wang2018lawn}.
Third, the byte-addressability of PMs facilitates the design of file systems. Specifically, PM-based systems can easily achieve lower write amplification facing small-grained write activities.
However, some challenges stay in the design of PM-based systems.
PMs have asymmetric read and write latencies, e.g., the writing activities in PCM can be 20$\times$ slower than reading.
Moreover, write activities shorten the lifetime of PMs.
Therefore, designs for PM-based systems should not impose overmuch writing activities on PMs.
Based on the above considerations, we review the existing lookup schemes for in-memory systems and propose our motivation.

\begin{figure}[tb]
    \centering
    \includegraphics[width=2.6 in,height = 2 in]{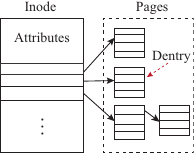}
    \caption{Conventional structure of directory files.}
    \label{fig:sequential}
\end{figure}

\subsection{Problems of Conventional Lookup Scheme}
Traditional structure of directory files is illustrated as Figure \ref{fig:sequential}.
When looking for a certain filename, the file system needs to sequentially seek dentries in a directory file.
This process takes $O(n)$ time to complete.
For block-based file systems, the performance of sequential reading is acceptable.

\begin{figure}[ht]
    \centering
\setlength{\abovecaptionskip}{0.2cm}
\setlength{\belowcaptionskip}{-0.2cm}
    \includegraphics[width=3.2 in,height = 1.1 in]{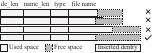}
    \caption{Conventional structure of dentries.}
    \label{fig:dlen}
\end{figure}

However, in PMFS, such conventional lookup schemes degrade the performance of file systems.
In particular, create operations magnify the defect in two ways.
First, it needs to traverse the whole directory to confirm if a file with the same name has been created.
Second, it needs to scan the \emph{d\_len} and \emph{name\_len} of each dentry to find a free space to insert the new one, as shown in Figure \ref{fig:dlen}.
Moreover, as for open operation, such one-by-one searching routines will be iteratively invoked during path lookup.
Above observation reveals that utilizing conventional lookup scheme cannot fully exploit the advantages of random read/write performance in the PM-based scenario.

\begin{figure}[ht]
    \centering
\setlength{\abovecaptionskip}{0.2cm}
\setlength{\belowcaptionskip}{-0.1cm}
    \includegraphics[width=2.8 in,height = 1.3 in]{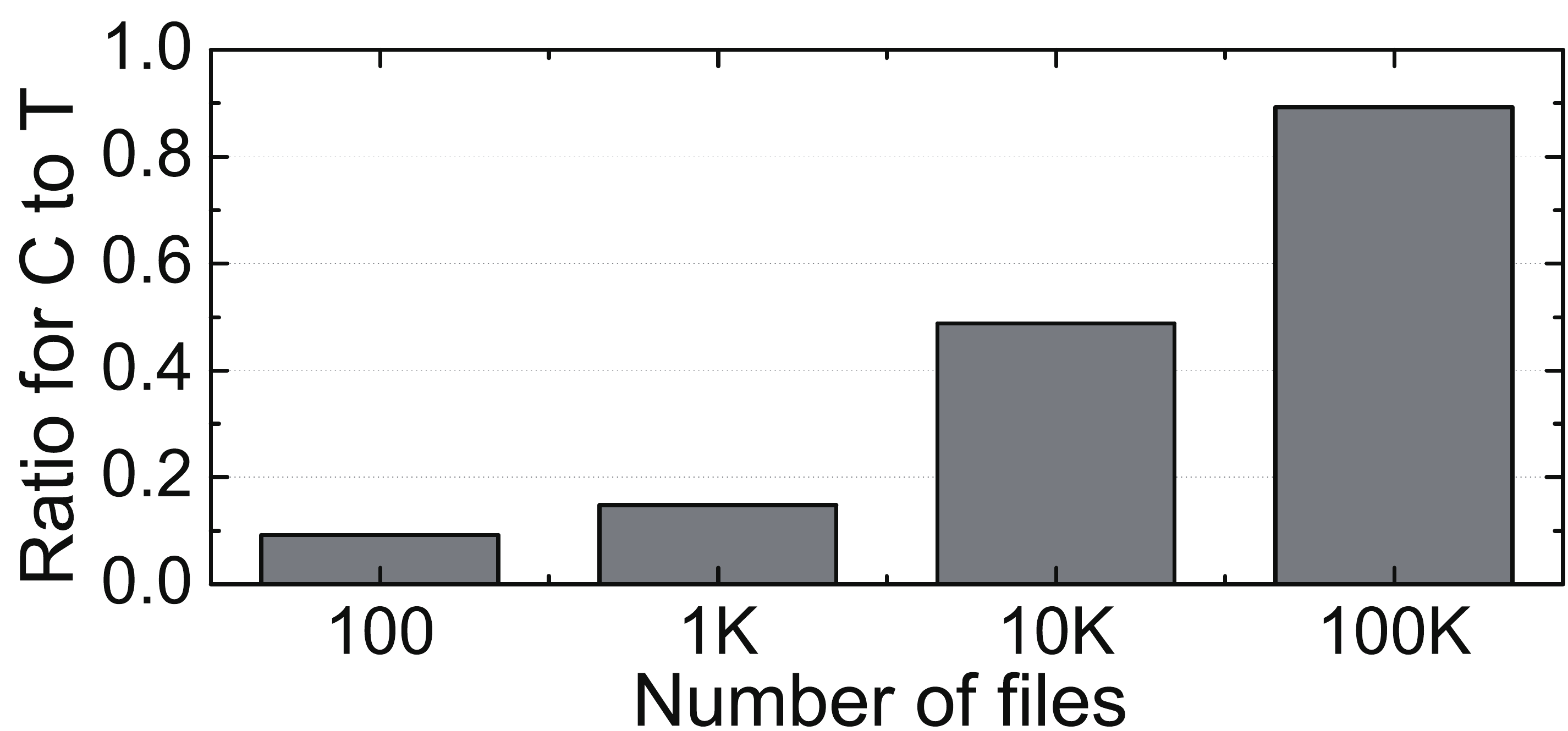}
    \caption{The proportion of time spent on directory operations.}
    \label{fig:ratio}
\end{figure}

We give a motivational example to demonstrate the inefficiency of conventional lookup scheme.
In our example, we first create 100 to 10K files in a directory, while writing 400K data for each file.
Then, we record the time of creating files $C$ and the total elapsed time $T$, respectively.
To put it plainly, $C$ represents the time spent on directory operations, while $T$ represents the total time employed by the file system.
Finally, we compare the ratio of $C$ to $T$.
The time $C$ scales exponentially with the increasing number of files.
This motivational example evidently illustrates the inefficiency of conventional lookup schemes.

\subsection{Problems of Extra Tree Indexes}
To improve the time performance of directory operations, recent schemes propose extra tree-based structures in DRAM as the index of directory files \cite{xu2016nova,rodeh2013btrfs}.
However, after power failures, they need to rebuild these extra structures to reboot the system.
As PMs provide persistent massive storage with DRAM-class speed, it is possible to put the auxiliary structures into PMs to accelerate the boot routine.

 \begin{figure}[t]
    \centering
\setlength{\abovecaptionskip}{0.3cm}
\setlength{\belowcaptionskip}{-0.1cm}
    \includegraphics[width=3.3 in,height=2.1 in]{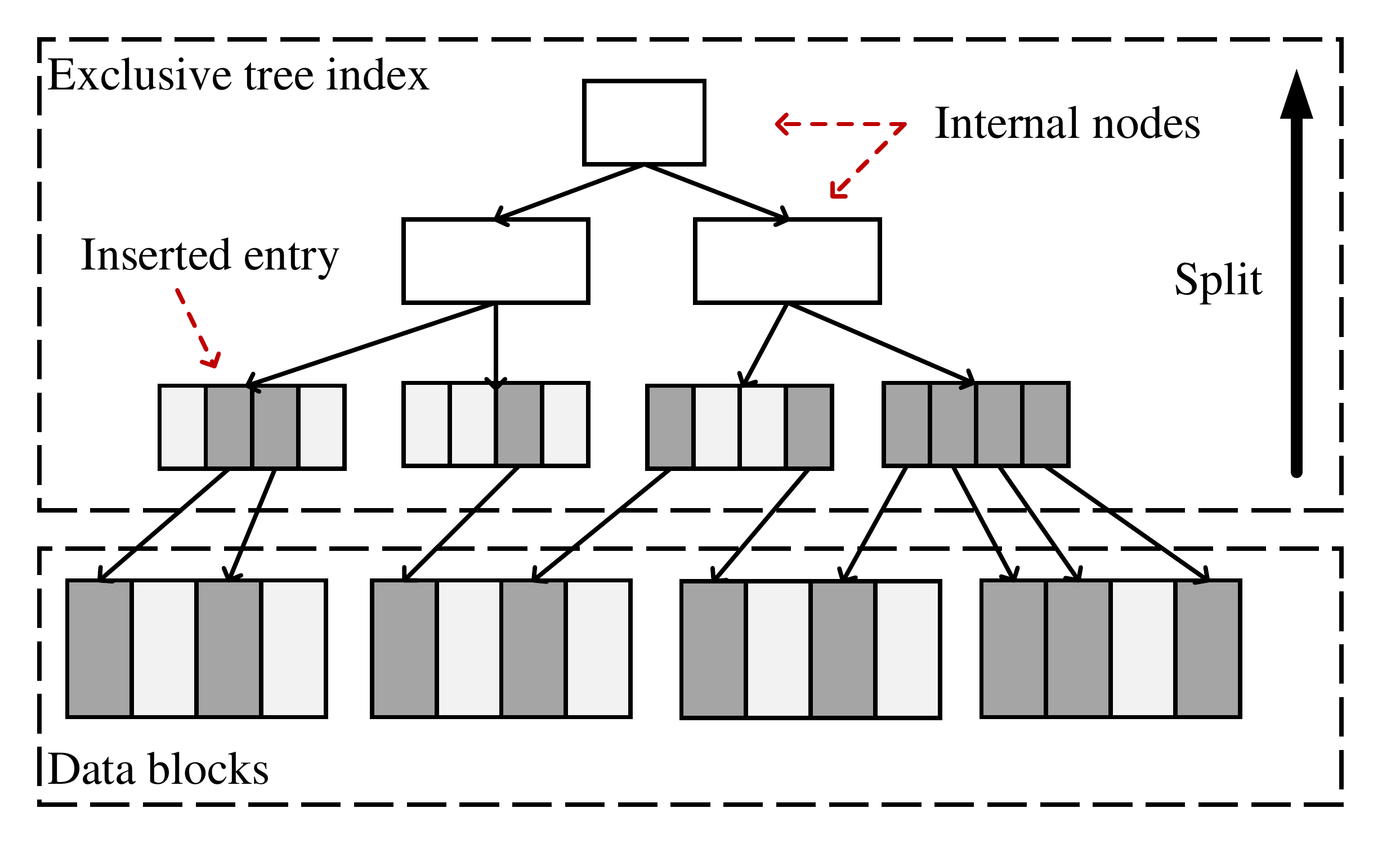}
    \caption{Extra tree index of dentries.}
    \label{fig:tree}
\end{figure}
However, such extra indexes are inadequate to be directly deployed in PMs.
Specifically, tree structures incur a large quantity of PM writes and massive memory consumption in three aspects.
\emph{First,} maintaining tree structure incurs a large quantity of PM writes.
When building tree indexes, filenames and other content of dentries will be inserted to leaf nodes.
This operation brings sorting of entries in traditional tree structures, which imports significant PM writes.
Insertions also trigger split operation when a leaf node is full, as shown in Figure \ref{fig:tree}.
Moreover, such split operation may be propagated to the root when its parent nodes are also full.
This chain effect imports considerable PM writes and therefore incurs degradation on performance and depresses the endurance of PMs.

\emph{Second,} for large-scale storages, maintaining tree indexes suffers from large memory consumption.
As mentioned above, content of dentries will be copied to leaf nodes, which nearly doubles the size of directory files.
What's more, massive internal nodes in tree structure exacerbate the consumption.
This problem lies in both DRAM and PM based systems.

\emph{Third,} on the consideration of consistency, tree structures usually adopt logging or shadowing mechanism, either of which aggravates the number of PM writes.
In another word, it sacrifices the performance of file system to maintain the validity of extra indexes.

Based on above analysis, it can be concluded that a novel lookup scheme is demanded to accelerate the directory operations with minimum overhead for PM-based file systems.
Our design principles of the lookup scheme for in-memory file systems are given as follows.
\begin{itemize}[noitemsep,topsep=0pt,parsep=0pt,partopsep=0pt]
  \item Operations on directory files should not traverse the whole field of dentries.
  \item On the consideration of memory consumption, the size of the extra structure should be minimized.
  \item The novel scheme should avoid too much consistency and maintenance overhead that will severely degrade system performance.
\end{itemize}

In the following two sections, we propose a novel lookup scheme CIB for in-memory file systems.

\section{CIB Design}\label{Sec:DesignFramework}

In the design of the lookup scheme, we need to resolve several problems.
First, how to redesign an elegant lookup routine to prevent sequential lookups.
Second, how to reconstruct directory files and reorganize dentries to facilitate the acceleration method.
Third, how to continually improve performance rather than introducing extra overhead.
This section will introduce the novel lookup scheme, ``Content-Indexed Browsing'' (CIB) to address the above problems.

\subsection{Regularize Dentries into fixed size}
\begin{figure}[thb]
    \centering
\setlength{\abovecaptionskip}{0.2cm}
\setlength{\belowcaptionskip}{-0.1cm}
    \includegraphics[width=2.3 in,height=0.6 in]{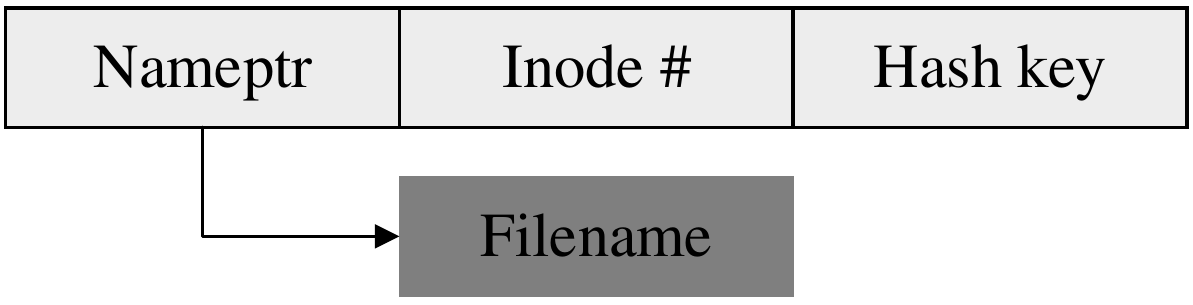}
    \caption{The structure of a dentry.}
    \label{fig:dentry}
\end{figure}

Motivated by section \ref{Sec:Motivation}, sequential lookup performs poor with the increasing data size of directory files.
Inspired by indexing schemes, we decide to aggregate dentries in terms of uniform keys rather than irregular filenames.
To make this possible, we redesign the structure of dentries.
As illustrated in Figure \ref{fig:dentry}, besides filename and inode number, we also store the hash key of filename.

To avoid modifying \emph{d\_len} and \emph{name\_len} in traditional schemes, we regularize the size of dentries to be fixed.
As the length of a filename is variable, we design a \emph{nameptr} to point to the real filename, so that we can use a pointer to represent a filename.
This structure of dentry fully leverages the byte-addressability of PMs.
Different from the traditional dentry searching, we first use a hash key to find the possible dentry, then we check its corresponding filename.
This mechanism utilizes uniform hash keys to reduce the overhead of comparing filenames.
Since comparing strings is much inefficient than comparing integer keys, CIB can significantly promote dentry lookups.

\subsection{Reconstruct Directory Files}
\begin{figure}[tb]
    \centering
\setlength{\abovecaptionskip}{0.1cm}
\setlength{\belowcaptionskip}{-0.1cm}
    \includegraphics[width=3.5 in,height = 1.1 in]{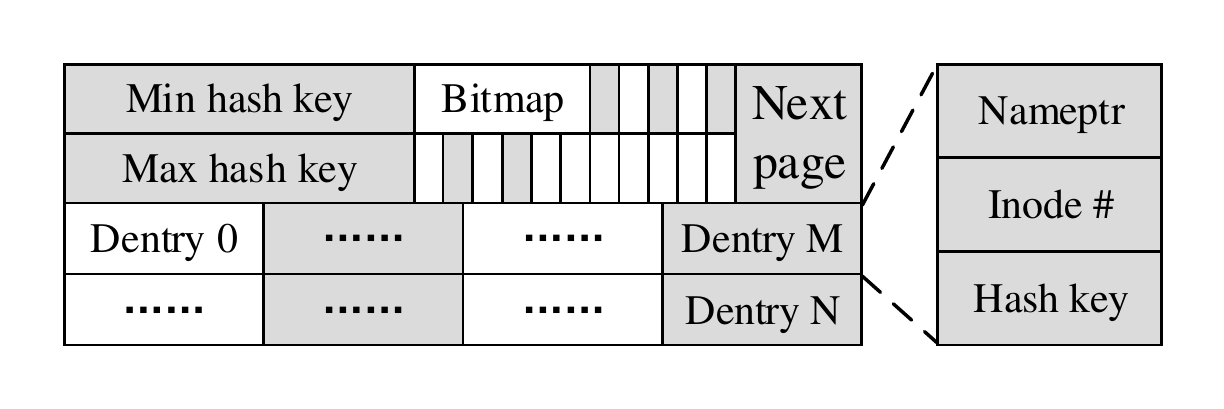}
    \caption{Reconstruction of directory blocks.}
    \label{fig:datapage}
\end{figure}

In PMFS, blocks of data are indexed by pointers stored in inodes.
This facilitates the random access of blocks for data files.
However, sequential search in directory files makes the index redundant.
Now, as we aggregate the dentries in terms of hashing keys, the structure of directory files needs to be totally redesigned.
In our scheme, the blocks of directory files are linked as a sorted list.
As shown in Figure \ref{fig:datapage}, each block has a minimum and maximum hash key to identify the range of keys stored in the block.
Each block contains dentries whose keys are within the range and the non-overlapping ranges of blocks in the list are relatively ordered.
Since the size of dentries is designed to be fixed, we can utilize a bitmap to indicate the state of each dentry.
Finally, the field \emph{next page} points to the following block in the linked list.

Based on the design, we mitigate the overhead for identifying dentries and maintaining the index.
Furthermore, we only need to store a single pointer in inode to access the head of the list, rather than store the whole index like traditional schemes.
And it is rather easy to guarantee the consistency of such linked lists without modifying inodes.

\subsection{Adaptive Acceleration Mechanism}
The directory files are frequently accessed and modified in file systems.
Accelerating the operations in terms of dentries is quite crucial for sustaining great performance of file systems.
Thus, we propose an adaptive acceleration mechanism to facilitate the access of blocks.
For directory files in small sizes, e.g., all dentries are stored in one block, sequential lookup for dentries has no side effect on performance.
As for directory files with a lengthy linked list of blocks, we build an auxiliary structure to improve its performance.

\begin{figure}[thb]
    \centering
\setlength{\abovecaptionskip}{0.2cm}
\setlength{\belowcaptionskip}{-0.1cm}
    \includegraphics[width=3 in, height=1.2 in]{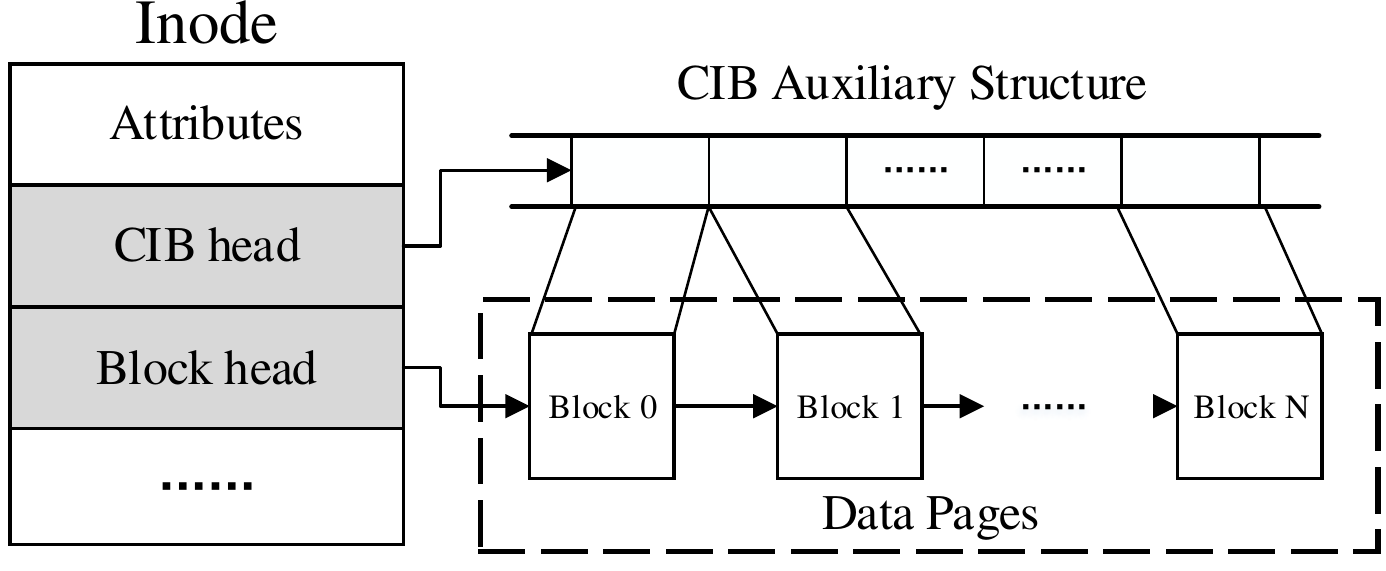}
    \caption{The overview layout of CIB.}
    \label{fig:blocklist}
\end{figure}

Inspired by high-performance binary search upon nodes in {B$^+$-tree}, we intend to make the scattered blocks ``gathered and flat''.
First of all, we apply an array of contiguous virtual addresses.
Then, we build a linear mapping between the array and the blocks.
As shown in Figure \ref{fig:blocklist}, we store the head address of the array into inode.
By this way, we can access the linked list of blocks randomly.
Thus, efficient searching algorithms such as binary search can be conducted upon our blocks.

It is worth noting that for the existing extra-indexing scheme, both the free space and inserted dentries in directory files need building exclusive indexes to acquire efficient management.
Otherwise, create operations still require sequential searches to find a segment of proper free space to insert new dentries.
On the contrary, CIB can efficiently manage the whole space of directory files by leveraging only a small array of addresses.
This is because CIB makes full use of the futures provided by reconstructed dentries and blocks.
Therefore, CIB significantly reduces the PM writes and memory consumption over extra index schemes.

By incorporating the futures of index and PMs, we reconstruct the dentries and directory files.
We abandon extra indexing schemes to avoid massive PM writes and design a new scheme to accelerate accessing the new structure.
Thus, the dentries are indexed with minimum overhead, which means, the directory files are \textbf{Content-Indexed}.
In the next section, we will detail the browsing routines of CIB.

\section{CIB implementation}\label{Sec:implementation}
This section first introduces the building routine of auxiliary structure.
Then we give the detailed basic routines for file operations based on our CIB.

\textbf{Building auxiliary structures.}
As the number of blocks in certain directory files grow up, we build an auxiliary array to accelerate locating blocks.
We first allocate a new array in virtual address space.
Then, we put the addresses of blocks into the array in order.
When accessing the cell in this array, we obtain the corresponding block.
By this way, we can randomly access dentry blocks through the array of continuous addresses.

\textbf{Open files.}
Open is the fundamental operation in file systems, since it is the pre-work of other operations, such as read and write operations.
Given the path of a certain file, the open operation is to lookup the path and find the inode number of the file.
As described in section \ref{Sec:DesignFramework}, we use \textbf{hash key} to filter out mismatched dentries with high efficiency, and use \textbf{filename} to confirm the dentry.
We conduct open operation as the following steps.
First, we hash the filename to obtain the key as the marker of the file.
Then, we conduct a binary search on the continuous virtual address using the hash key.
After locating the block that may contain the target dentry, we start searching inside the block.
When encountering a dentry with the same key, we will check its corresponding bit in the bitmap of the block to identify the state of dentry.
If the bit equals 1, indicating that this dentry is valid, we continue to check the filename of this dentry to confirm whether it is the target dentry.
Finally, we get the inode number of the target file or return an error code if the file doesn't exist.

\textbf{Create files.}
To create a new file, its parent directory file needs to insert a new dentry.
First of all, we hash the filename and make the new dentry.
Based on the open operation, we can easily check whether there is another file with a duplicate filename.
If not, we find a free position in the block according to the bitmap.
Then we insert the new dentry into this position and set the corresponding bitmap to 1.

Note that once the block is full of dentries, we will create a new block to amortize the dentries.
First, we divide dentries into two sets in terms of their hash keys.
Then we allocate a new block and modify the ''min'' and ''max'' hash keys.
We copy (not move, for consideration of arbitrary system failures) the set of dentries with larger keys into the new block.
Finally, we link the new bucket into the block-list just after the original one.

\textbf{Delete files.}
Different from the complex deletion routine of traditional schemes that will modify several variables in inodes, delete operation in CIB is rather concise.
Same as creation operation, we first locate the target dentry in terms of the key and filename.
After that, we atomically modify the corresponding bit in the bitmap to 0 to make this dentry invalid.
That is to say, we complete deletion of dentry with only 1 bit PM writes.

\textbf{Consistency and recovery.}
In CIB, most operations are atomically conducted without recording logs.
For building operation, the head of contiguous addresses is stored in the inode as a pointer.
Since PMs can provide atomic 8-byte writes, the modification of this pointer will not introduce inconsistency.
Similarly, we utilize this future to guarantee the consistency of creation and deletion operations.
Specifically, we manage the space of blocks and identify the validation of dentries using bitmaps, which can be flushed into PMs atomically.

As for applying a new block to amortize dentries, we set a flag in the bitmap to mark the status of the two affected blocks.
When a system failure occurs, there may be two inconsistent situations.
If the original one fails to link the newly created block, the whole list is still available after reboot.
If the new block has been inserted into the list while duplicate dentries in the original one are not invalidated yet, the flag in the bitmap will remind the system fixing this problem.
By this way, CIB costs minimum overhead for consistency and requires miniature recovery routines after system failures.

\section{Experimental Results}\label{Sec:Exp}

We implement the ``Content-Indexed Browsing'' ({CIB}) mechanism and the redesigned directory structure in PMFS \cite{dulloor2014system}, denoted as CIB-PMFS.
Evaluation experiments are conducted on a workstation equipped with 256GB main memory and two Intel(R) Xeon(R) E5-2640 processors.
We partition 10 GB DRAM to stand for the PM device used for PMFS.
We compare {CIB-PMFS} with PMFS in terms of time performance for directory-related operations.
The operating system is Ubuntu 16.04, with Linux kernel 4.4.4.
We evaluate the performance and writes of CIB via the widely-used benchmark Filebench \cite{tarasov2016filebench}.

\subsection{Performance Evaluation}
This subsection first evaluates the time performance of CIB and conventional lookup scheme of PMFS, then demonstrates the overall performance of CIB-PMFS and original PMFS.
We vary the number of files manipulated in each workload and record the elapsed time of directory operations.
Specifically, we first create $10^{3}$ to $10^{5}$ files to perform the loading phase of directory files.
Then we utilize other workloads to further demonstrate the efficiency of the CIB lookup scheme.
After that, we integrate CIB into PMFS to evaluate the overall performance of the file system.
Finally, we give a comparison between CIB and extra tree structure to prove the PM-friendliness of CIB.

\begin{table}[htbp]
  \centering
\setlength{\abovecaptionskip}{0.2cm}
\setlength{\belowcaptionskip}{-0.1cm}
  \caption{Elapsed time for inserting dentries (sec).}
    \begin{tabular}{|l|r |r |r |r |}
    \hline
    \centering
    \# of files   & 1K  & 10K & 50K & 100K\\
    \hline
    Trad.   & 0.017209 & 0.946182 &16.462971& 45.54016\\
    \hline
    CIB    & 0.003566 & 0.036723 & 0.16732&0.258627\\
    \hline
    \textbf{speedup}   & \textbf{4.83$\times$}  & \textbf{25.77$\times$}& \textbf{98.39$\times$}  & \textbf{176.08$\times$}  \\
    \hline
    \end{tabular}%
  \label{tab:Exp_loading}%
\end{table}%
Table \ref{tab:Exp_loading} reports the elapsed time of two schemes in creating phase.
In this table, line \emph{\# of files} represents the number of created files, while lines \emph{Trad.} and \emph{CIB} give the elapsed time of the traditional lookup scheme in PMFS and the proposed CIB respectively.
We have three observations for this result:
(1) CIB speedups over the conventional lookup scheme from 1.07$\times$ to 176.08$\times$.
(2) The elapsed time of the conventional lookup scheme scales up quickly as analyzed in section \ref{Sec:Motivation}.
Especially, when the number of files increases from $10^{4}$ to $10^{5}$, the elapsed time increases more than 48$\times$.
There are two reasons that cause performance degradation.
First, PMFS adopts inefficient sequential lookup scheme.
Second, PMFS compares every filename in the lookup routine, while string comparison costs a lot.
(3) CIB achieves stable performance in any magnitudes of files.
When the number of files is small, the increases of elapsed time are roughly the same with that of files.
However, the elapsed time for inserting $10^{5}$ dentries is only 1.65$\times$ to that for inserting $10^{6}$ dentries.
This is because CIB utilizes uniform hash keys to conduct binary search and thus dramatically reduce lookup overhead.

Based on various numbers of inserted dentries, we further evaluate CIB with traces generated by Filebench.
We choose two workloads \emph{Webproxy} and \emph{Varmail} to perform operations on two lookup schemes.
It can be observed that CIB significantly outperforms the traditional lookup scheme in dir-sensitive workloads.

\begin{figure}[htbp]
    \centering
\setlength{\abovecaptionskip}{0.3cm}
\setlength{\belowcaptionskip}{-0.1cm}
    \includegraphics[width=3 in,height=1.35 in]{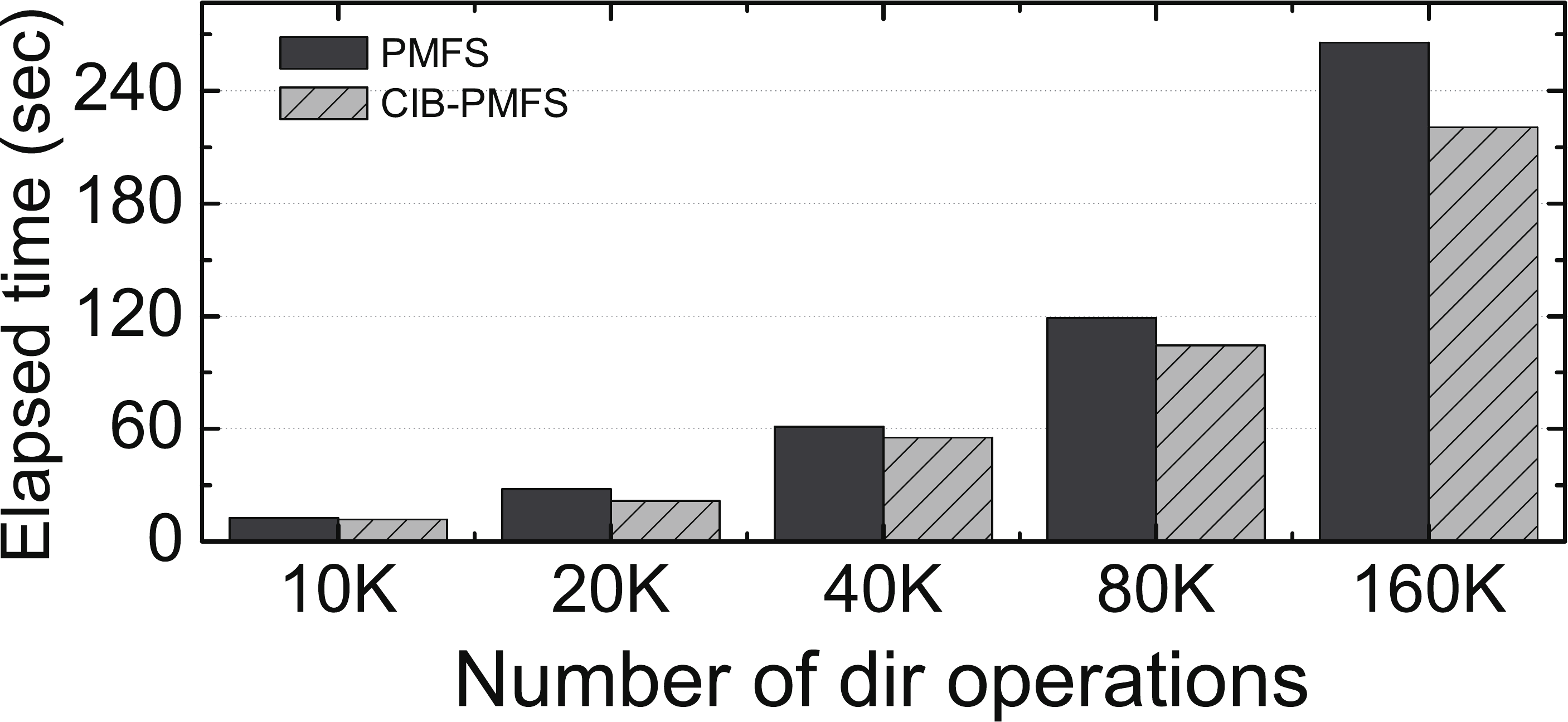}
    \caption{Comparison of overall elapsed time.}
    \label{fig:Exp_create}
\end{figure}
Moreover, we evaluate the \textbf{overall} performance of the file system with various directory-related operations generated by Filebench Create workload.
It is noteworthy that, although overall evaluations create massive tiny-scale directory files and suffer transactional overheads, CIB-PMFS still improves the overall performance steadily.
The proposed CIB improves the overall performance of the file system by 8\% to 20.4\%.
This is because the CIB mechanism accelerates path lookup routine and thus facilitates all other file operations.
These experiments highlight the effectiveness of CIB with real-world traces.

\subsection{Comparison with Extra Index}
As BtrFS \cite{rodeh2013btrfs} and NOVA \cite{xu2016nova} utilize tree structures for lookup routines, we compare CIB with {B$^+$-tree} to prove the PM-friendliness of CIB.
We insert various numbers of dentries into each of them and record the number of PM writes.
Comparing with {extra tree indexes}, {CIB} mechanism reduces PM writes by orders of magnitude.
This is because CIB takes full advantage of the directory structure without building massive extra index structures.

Additionally, we compare the timing performance of inserting 50M, 60M, and 100M dentries, respectively.
CIB steadily outperforms extra index schemes in this evaluation.
Specifically, the elapsed times of CIB achieve 14.1$\times$, 15.6$\times$, and 20.2$\times$ reduction over the original {B$^+$-tree} for each workload.
It can be attributed to two reasons: 1) CIB conducts less string comparisons and structural changes; 2) CIB leverages a concise structure and thus mitigates consistent overhead.
This experiment proves that CIB is an efficient PM-friendly lookup scheme.

\section{Conclusion}\label{Sec:Con}
This paper investigates the existing lookup schemes from the perspective of PM-based in-memory systems.
We reconstruct the dentries and data blocks to mitigate redundant overhead.
Then we present a novel lookup scheme, Content-Indexed Browsing (CIB) to achieve significant performance improvement.
This scheme helps in-memory file systems further exploit the potential of PMs and prevent numerous PM writes.
We implement {CIB} in PMFS and conduct extensive evaluations to illustrate the efficiency of it.
Results show that {CIB} achieves times of speedup over the PMFS and significantly reduces the number of PM writes over extra index schemes.

The major point of this paper is to incorporate the merits of indexes into directory files to accelerate path lookups for PM-based file systems.
By reconstructing the dentries and directory blocks, it can retrieve the content of directory files efficiently without maintaining exclusive indexing structures.
In other words, the auxiliary structure implemented in this paper is not the unique and fixed matching for CIB.
Thus, the CIB provides the opportunity to be customized with other types of indexes (such as trees and hashing tables) orienting to application requirements.

\bibliographystyle{IEEEtrans}
\bibliography{sample-bibliography}

\end{document}